\documentclass[preprint,12pt]{elsarticle}

\usepackage{amssymb}
\usepackage{amsmath}

\journal{International Journal of Multiphase Flow}

\begin{document}

\begin{frontmatter}



\title{Effect of converging shape of container on the velocity of impact-induced focused liquid jet} 

\author[inst1]{Hiroya Watanabe} 

\affiliation[inst1]{organization={Department of Mechanical Systems Engineering, Tokyo University of Agriculture and Technology},
            addressline={Nakacho 2-24-16}, 
            city={Koganei},
            postcode={184-8588}, 
            state={Tokyo},
            country={Japan}}

\author[inst2]{Hiroaki Kusuno} 
\affiliation[inst2]{organization={Department of Mechanical Engineering, Faculty of Engineering Science, Kansai University},
            addressline={Yamate-cho 3-3-35}, 
            city={Suita},
            postcode={564-8680}, 
            state={Osaka},
            country={Japan}}
\author[inst1]{Yoshiyuki Tagawa} 

\begin{abstract}
We investigated the effect of container shape on the behavior of the impact-induced focused liquid jets by dropping a converging-shaped container (e.g., Kjeldahl flask) partially filled with liquid onto a floor to develop a method for increasing the jet velocity. 
Note that similar well-known experiment, Pokrovski's experiment, in which a focused liquid jet is generated by the impact of a cylindrical test tube is free-from the effect of the converging shape.
The results showed that the jet was up to about 1.6 times faster in a converging-shaped container than in a test tube, despite the same impact velocity.
To understand the mechanism of the increase in the jet velocity, the Laplace equation on the pressure impulse was solved numerically under the boundary condition that the pressure impulse is given at the bottom of the container.
The pressure impulse field was then obtained.
The normalized gas-liquid interfacial velocity obtained from the numerical solution of the pressure impulse field agrees well with the normalized jet velocity in experiments, showing that impact-induced focused liquid jets are driven by the pressure impulse generated at the bottom.
In addition, numerical solutions of pressure impulse fields in a simpler shaped container with different degrees of convergence were compared with analytical solutions obtained from a lower order model of the pressure impulse field. 
We confirmed that the gas-liquid interfacial velocity of the impact-induced focused liquid jet is governed by changes in both the flow rate and pressure impulse gradient at the central axis of the container caused by changes in the cross-sectional area of the container.
We showed that by changing the container shape, we can increase the velocity of the gas-liquid interface after the container impact.
This finding is expected to be applied to the ejection and application of high-viscosity liquids such as paints, sealants, and adhesives without dilution, as well as to needle-free injection technology using fast focused liquid jets.
\end{abstract}




\end{frontmatter}



\section{Introduction}
\label{Intro}
Rapid changes in liquid velocity due to impact and the accompanying violent motions at the gas-liquid interface have been studied for many years.
As an example, to understand the phenomena of wave crashing against a seawall, the pressures and loads acting on the vertical wall (\citet{Bagnold1939-ly, Hattori1994-vl, Cooker1995-vr, Cuomo2010-uh,Altomare2015-hr}) and the intense gas-liquid interfacial motion(\citet{Cooker1992-wa, Bredmose2015-jj}) have been studied.
The other impact phenomena of solids and liquids, such as droplet impact on substrates (\citet{Lesser1983-vl, Yarin2006-br, Josserand2016-dd, Wu2021-tv, Cheng2022-sk, Gonzalez-Avila2024-io}), slamming of ships (\citet{Dias2018-jh}), and the water entry of solid objects (\citet{Aristoff2009-aa, Truscott2014-or, Stavropoulos-Vasilakis2019-ma}), have also been widely studied.

The sudden change in motion of liquid in a container due to impact is of great engineering importance in terms of mixing, transport, and storage of liquids.
Therefore, many studies have been conducted on the occurrence of cavitation associated with impact (\citet{Pan2017-sx}) and the associated phenomena of jetting, erosion, and destruction (\citet{Daily2014-ki, Blake2015-rc, Supponen2016-on}).

Pokrovski's experiment is known as a phenomenon in which container impact causes dynamic motion at the gas-liquid interface.
In Pokrovski's experiment, a glass or metal test tube filled with liquid was dropped onto a table.
The impact induced the generation of a focused thin liquid jet from the gas-liquid interface (\citet{Lavrentiev1980-or})(See a supplementary movie 1).
In the generation of such an impact-induced focused liquid jet, the rapid velocity change due to container impact and the flow focusing at the concave-shaped gas-liquid interface are important (\citet{Antkowiak2007-ax}).
Similar jets have been reported to be generated in systems where the container comes to a sudden stop without bouncing, indicating a large acceleration change given to the entire system of container and liquid is important (\citet{Krishnan2022-wd}).

Since the impact-induced focused liquid jet extends straightly with a thin fast tip, it is expected to be applied as a method of liquid ejection and precise application.
The control of the jet velocity is essential for practical use of the liquid jet.
Velocity changes in liquids caused by sudden boundary motion, such as an impact of the container, can be described by the pressure impulse theory (\citet{Batchelor1967-tv, Cooker1995-vr}).
\citet{Kiyama2014-bh} reported a semi-empirical equation, which is based on pressure impulse theory and flow focusing effects (\citet{Tagawa2012-ci, Peters2013-xk}), can well describe the velocity of impact-induced focused liquid jets with a glass test tube.
With a stronger impact of the test tube, cavitation can be observed, which increases the jet velocity (\citet{Kiyama2016-tf}).
In the presence of cavitation, the propagation of pressure waves caused by the impact is attenuated, so the pressure impulse theory, which assumes incompressibility, cannot explain the mechanism of the increase in jet velocity.
\citet{Onuki2018-xi} inserted a small tube inside a test tube and pushed down the position of the gas-liquid interface inside the inner tube relative to the gas-liquid interface position outside the inner tube.
The difference in liquid level between the inside and outside of the inner tube increases the pressure impulse gradient inside the inner tube, resulting in successful ejection of a liquid with a kinematic viscosity of 500 $\mathrm{mm/s^2}$.
\citet{Kamamoto2021-xv} used a syringe-shaped ejection device that has a short nozzle and does not clog easily, and succeeded in the downward ejection of car body paint without dilution.
\citet{Kobayashi-2024-aw} have succeeded in the continuous ejection of silicone oil with a kinematic viscosity of 500 $\mathrm{mm/s^2}$ at an ejection frequency of 10 Hz with the syringe-shaped ejection device.
To increase the ejectable liquid viscosity for the industrial application, a mechanism is needed that increases the velocity of the liquid jet.

In this study, we investigate the impact-induced liquid jet phenomenon in a converging-shaped container to obtain knowledge other than cavitation effects that contribute to the increase in the jet velocity.
The sudden change in liquid velocity caused merely by the impulsive motion can be explained by the pressure impulse theory.
The pressure impulse field is obtained by solving the Laplace equation theoretically or numerically.
At this point, appropriate boundary conditions for the pressure impulse itself or its gradient, such as the Dirichlet and the Neumann conditions, respectively, are required.
In a previous study, the pressure impulse distribution for a test tube of infinite length was acquired by approximately solving the combined general solution of the Laplace equation of polar and cylindrical coordinates (\citet{Antkowiak2007-ax}).
In the case of a container with no cross-sectional area change, the same solution is obtained whether the boundary condition at the bottom of the container is the Direcret or the Neumann condition.
On the other hand, for the container with changing cross sections, which is the subject of this study, the Dirichlet and the Neumann conditions at the bottom have different meanings.

We experimentally investigated the effect of the converging shape of the container on the jet velocity by generating jets with such a container, where the cross-sectional area decreases from near the bottom to the mouth of the container. 
The experiments were performed under the situation where cavitation does not occur (or its effect on the jet velocity is limited).
As a result, faster jets were generated in a converging container (e.g., Kjeldahl flask) than in a container with no cross-sectional area change.
To understand the mechanism of the phenomenon, the Laplace equation on the pressure impulse is solved numerically under the experimental container shape.
From the numerically solved pressure impulse field, the liquid velocity at the bottom of the gas-liquid interface is calculated and compared with the experimentally measured jet velocities.
Next, to confirm the effect of the converging shape, we numerically investigate the pressure impulse field in containers with different degrees of convergence.
In addition, we consider a one-dimensional idealized model of the pressure impulse field in the converging shape container for the essential understanding of the change in the pressure impulse field due to the converging shape. 

The paper is organized as follows. Section \ref{Theory} describes the pressure impulse theory; Section \ref{Methods} outlines the experiments and the numerical setup for solving the Laplace's; Section \ref{Results and discussion} presents the experimental results and discusses the mechanism through comparison with numerical solutions; Section \ref{Conclusion} is the conclusion.

\section{Pressure impulse theory}
\label{Theory}
This chapter describes the pressure impulse theory that governs the liquid velocity change caused by the impulsive motion.
The jet velocity of interest in our research is measured at O(10) ms after the container's impact.
In contrast, the timescale for the propagation and reflection of pressure waves generated at the bottom of the container by impact is O($10^{-1}$) ms, which means that the jet velocity is determined by a well-developed pressure field.
Thus, we consider the velocity field of a liquid based on the assumption of incompressibility, where the pressure field is instantaneously generated after the container's impact.
In this case, the liquid velocity change associated with the impulsive motion of the liquid boundary can be approximated by only the time and pressure terms by comparing the order of each term from the incompressible NS equation.
This is because the impact duration is very short and the time derivative term is sufficiently large compared to the external force term, viscosity term, etc.
By integrating this equation over the duration $\tau$ of the impulsive motion, the difference between the speed immediately before the change $\boldsymbol u'$ and the speed immediately after the change $\boldsymbol u''$ can be expressed with the pressure impulse $\Pi$ using the following equation.

\begin{equation}\label{Eq:pressure-impulse}
\boldsymbol u'' -\boldsymbol u'= -\frac{1}{\rho} \nabla \Pi,~~~~~
\Pi = \int^\tau_0 p dt,
\end{equation}
where $\rho$ is the density and $p$ is the pressure of liquid. 
Eq. \ref{Eq:pressure-impulse} shows that the velocity change of the liquid due to impulsive motion is determined solely by the density and the gradient of the pressure impulse, without the influence of viscosity, surface tension, or gravity.
From Eq. \ref{Eq:pressure-impulse} and the continuity equation $\nabla \cdot u = 0$ for an incompressible fluid, the Laplace equation for the pressure impulse is derived.

\begin{equation}\label{Eq:Laplace}
\nabla^2 \Pi = 0.
\end{equation}
By solving Eq. \ref{Eq:Laplace} theoretically or numerically with the boundary conditions, the pressure impulse field can be obtained. For example, the pressure impulse field in an infinitely long test tube after impact with the floor, which induces the formation of a focused liquid jet, is approximately solved by a combination of general theoretical solutions of the Laplace equation in polar and cylindrical coordinate representations and boundary conditions (\citet{Antkowiak2007-ax}).

Based on previous studies (\citet{Batchelor1967-tv, Cooker1995-vr, Antkowiak2007-ax}), we describe how to provide boundary conditions for solving the Laplace equation to obtain the pressure impulse field in the case of the impact-induced focused liquid jet.
First, the pressure impulse $\Pi$ is zero at the gas-liquid interface since the pressure can be regarded as atmospheric pressure.

\begin{equation}\label{Eq:theory-B1}
\Pi=0,
\end{equation}
Here, we consider the case of a container falling with velocity $U_0$ until its velocity becomes zero due to impact.
At this time, the motion of the free-falling liquid according to gravity is suddenly stopped by the bottom of the container, the region where the inward normal vector of the wall has an upward component.
Thus, high pressure is generated in this region, which in turn generates a high pressure impulse.
The pressure impulse acting at the infinitely deep bottom can be expressed as

\begin{equation}\label{Eq:theory-B2}
\Pi=-\rho U_0 z,
\end{equation}
referring to a previous study (\citet{Antkowiak2007-ax}).
The $z$-axis is vertically upward, and $z=0$ is the bottom of the gas-liquid interface.
When an impulsive motion of a container causes a velocity change of the liquid inside, it may seem reasonable to give the Neumann condition to set the liquid velocity at the bottom of the container equivalent to the container velocity.
We will compare the pressure impulse fields solved numerically for a container of simple shape with a flat bottom, giving the Dirichlet and the Neumann boundary conditions at the bottom in \ref{Results-Sub3}.
The remaining liquid boundary is the wall condition, where no flow enters or exits.

\begin{equation}\label{Eq:theory-B3}
\frac{\partial \Pi}{\partial \boldsymbol n}=0,
\end{equation}
$\boldsymbol n$ is the inward normal vector of the wall.
By solving the Laplace equation on the pressure impulse together with the boundary conditions above, the pressure impulse field is calculated, and the velocity difference before and after the impulsive motion can be obtained from Eq.\ref{Eq:pressure-impulse}.

\section{Methods}
\label{Methods}

\subsection{Experimental setup}
\label{Methods-Sub1}
\begin{figure}
    \centering
    \includegraphics[width=0.7\linewidth]{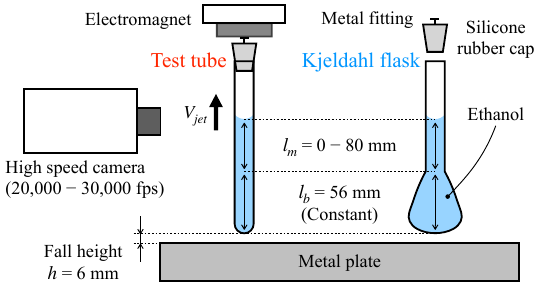}
    \caption{Schematic of the experimental setup.}
    \label{fig:Setup}
\end{figure}

To investigate the effect of the container's converging shape on the velocity and behavior of impact-induced focused liquid jets, we performed jet generation experiments using two types of containers at several liquid heights.
Fig. \ref{fig:Setup} shows the experimental setup.
We used a test tube (IWAKI, TEST15NP, inner diameter: 12.4 mm) and a Kjeldahl flask (IWAKI, 5420FK30, inner diameter of neck: 11.9 mm) to generate impact-induced focused liquid jets.
Since a concave shaped gas-liquid interface is essential for the generation of focused liquid jets (\citet{Antkowiak2007-ax}), hydrophilic treatment of the inner walls of containers was performed as a preliminary preparation.
First, a 10 wt$\%$ sodium hydroxide solution was generated by dissolving sodium hydroxide (Hayashi Pure Chemical Industry, special grade pearlescent 19002485) in pure water (Merck, Milli-Q), the generated solution was placed in containers and kept for 30 minutes, then the containers were washed with pure water (Merck, Milli-Q) and dried.
In this experiment, the shape of the gas-liquid interface did not change significantly for each condition.
Both containers were partially filled with ethanol (Hayashi Pure Chemical Industry, Ethanol 99.5 $\%$), capped with a silicone rubber cap (Aram, No. 1) with a screw attached.
To avoid the occurrence of cavitation, we dropped containers from a drop height of 6 mm onto a metal floor (Toshin Steel, material: SS400, thickness $\times$ width $\times$ length: 20 $\times$ 150 $\times$ 150 mm) using an electromagnet (Fujita, round electromagnet FSGP-40). 
In all experiments, the bottom part of both containers ($l_\mathrm{b}$ part in the Fig.\ref{fig:Setup}) was filled with ethanol. 

The main parameter except the shape of containers in this experiment is the height of the liquid column $l_\mathrm{m}$ in the neck part of the Kjeldahl flask or in the middle part of the test tube (see Fig.\ref{fig:Setup}).
We then observed the behavior of liquid jets with varying $l_\mathrm{m}$ from 0 to 80 mm.
Note that, although the weight of the whole container-liquid system changed by varying the height of the liquid in the container, the change in velocity given to the system remained constant since the drop height was fixed for all conditions.

The images were taken with a high-speed camera (Photron, FASTCAM SA-X2), and used for the observation of the jet behavior and the measurement of the jet velocity.
The frame late of camera is 20,000-30,000 f.p.s., and the resolution is 0.22 mm/pixel.
Since the cavitation bubble formation was not observed, we consider the effect of cavitation on the jet velocity is negligible in this experiment.
In this experiment, the liquid jet was generated three times for each liquid height 
condition.
The displacement of the jet tip relative to the container position between 7 ms and 15 ms after the time of container impact was analyzed from obtained images.
Then the jet velocity $V_\mathrm{jet}$ was calculated by dividing the displacement by the time required for the displacement ($\Delta t =8 \mathrm{~ms}$).
The mean and the standard deviation of the jet velocity in three trials in each liquid height condition are compared in \ref{Results-Sub1}.

\subsection{Numerical solution of the Laplace equation}
\label{Methods-Sub2}
This section describes how to solve the Laplace equation numerically to obtain the pressure impulse generated by container's impact in the test tube and the Kjeldahl flask.
Using the commercial software COMSOL Multiphysics, the Laplace equation (Eq.\ref{Eq:Laplace}) is expressed in cylindrical coordinates assuming axial symmetry as follows,

\begin{equation}\label{Eq:Laplace-cylindrical}
\frac{\partial^2 \Pi}{\partial r^2}+\frac1r\frac{\partial \Pi}{\partial r}+\frac{\partial^2 \Pi}{\partial z^2}= 0.
\end{equation}
We set the diameter of the liquid column in the test tube to 12.4 mm, and the the bottom shape was set to be a perfect hemisphere.
The shape of the Kjeldahl flask was reproduced from the image taken with a high-speed camera.
The diameter of the liquid column at the neck of the Kjeldahl flask was set to 11.9 mm.
The shape of the gas-liquid interface in the numerical analysis was set to be a perfectly hemispherical with the contact angle of 0° for both containers, regardless of the liquid height in the middle or neck part $l_\mathrm{m}$.

\begin{figure}
    \centering
    \includegraphics[width=0.9\linewidth]{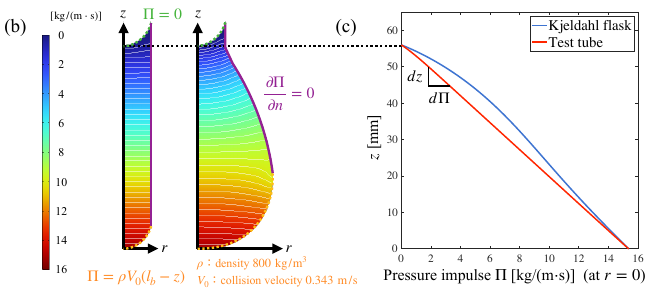}
    \caption{(a) Boundary conditions of numerical analysis and the calculated pressure impulse field in each container. (b) The pressure impulse distribution on the central axis of each container. The analysis was conducted on the condition of $l_\mathrm{m}=0$ mm. The velocity at the gas-liquid interface was calculated from the pressure impulse gradient at the bottom of the gas-liquid interface and compared with the experimentally obtained jet velocity in \ref{Results-Sub2}.}
    \label{fig:Pi-field}
\end{figure}

We solved Eq.\ref{Eq:Laplace-cylindrical} numerically with the boundary conditions.
The case with the condition of $l_mathrm{m}=0$ mm is shown in Fig. \ref{fig:Pi-field}(a) as a representative example.
At the boundary corresponding to the gas-liquid interface (see Fig. \ref{fig:Pi-field}(a) : green dashed line), the pressure can be regarded as atmospheric pressure ($p=0$), and thus $\Pi=0$.
At the bottom of the container, the region where the $z$-axis component of the inward normal vector $\boldsymbol n$ has a positive component (Fig. \ref{fig:Pi-field} : yellow dashed line), the following Dirichlet condition is given as described in Chapter \ref{Theory}.
$\Pi=\rho U_0 (l_\mathrm{b}+l_\mathrm{m}-z)$.
Here, the liquid density $\rho$ is 800 $\mathrm{kg/m^3}$ and the initial velocity $U_0$ is 0.343 m/s.
The remaining boundaries (see Fig. \ref{fig:Pi-field} : purple solid line) are subjected to the zero flux condition, which suppresses the entry and exit of the flow from the wall.
$\partial \Pi / \partial \boldsymbol n =0$.

\section{Results and discussion}
\label{Results and discussion}

\subsection{Experimental results}
\label{Results-Sub1}
In this section, we describe how the kinematics of impact-induced focused liquid jets change when the height of the liquid in the neck is varied in a container with a converging shaped container, a Kjeldahl flask, compared to a test tube with no cross-sectional area change.

\begin{figure}
    \centering
    \includegraphics[width=1.0\linewidth]{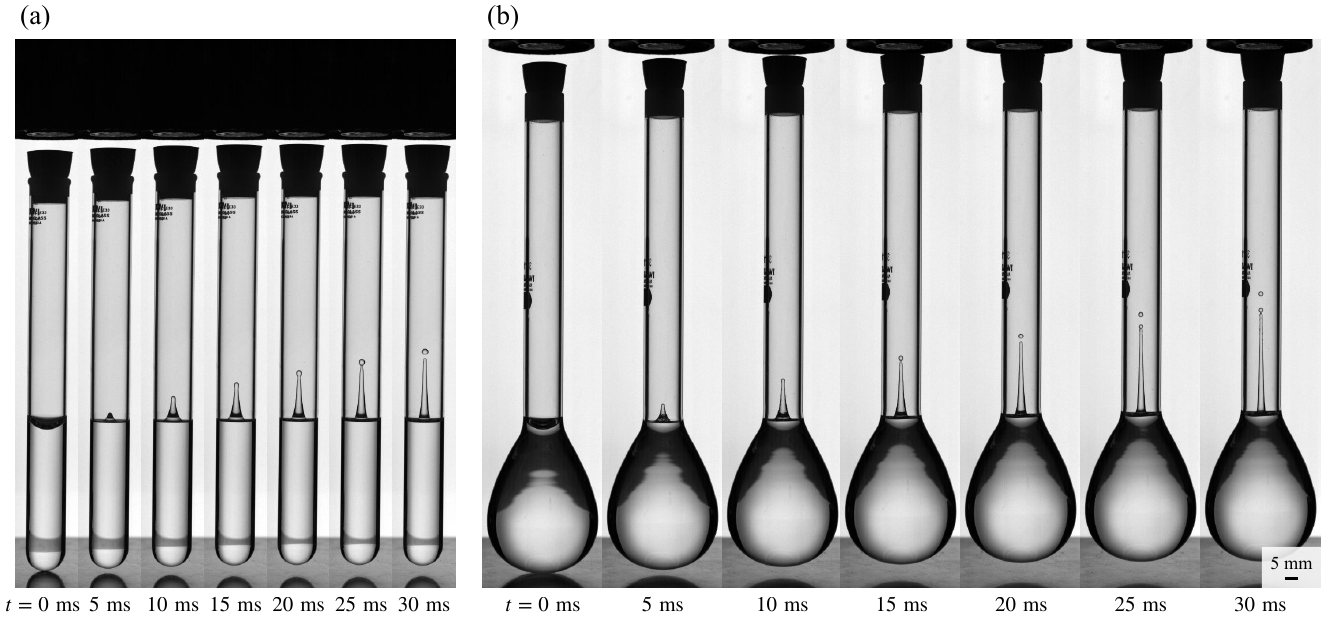}
    \caption{Impact-induced focused liquid jets with (a) the test tube and (b) the Kjeldahl flask. A test tube and a Kjeldahl flask partially filled with ethanol ($l_\mathrm{m}=0$ mm) were dropped from a height of 6 mm. In both containers, upon container collision, a focused liquid jet was generated and elongated to produce a droplet in a similar manner. The time interval between each image is 5 ms.}
    \label{fig:Jet-time}
\end{figure}

\begin{figure}
    \centering
    \includegraphics[width=1.0\linewidth]{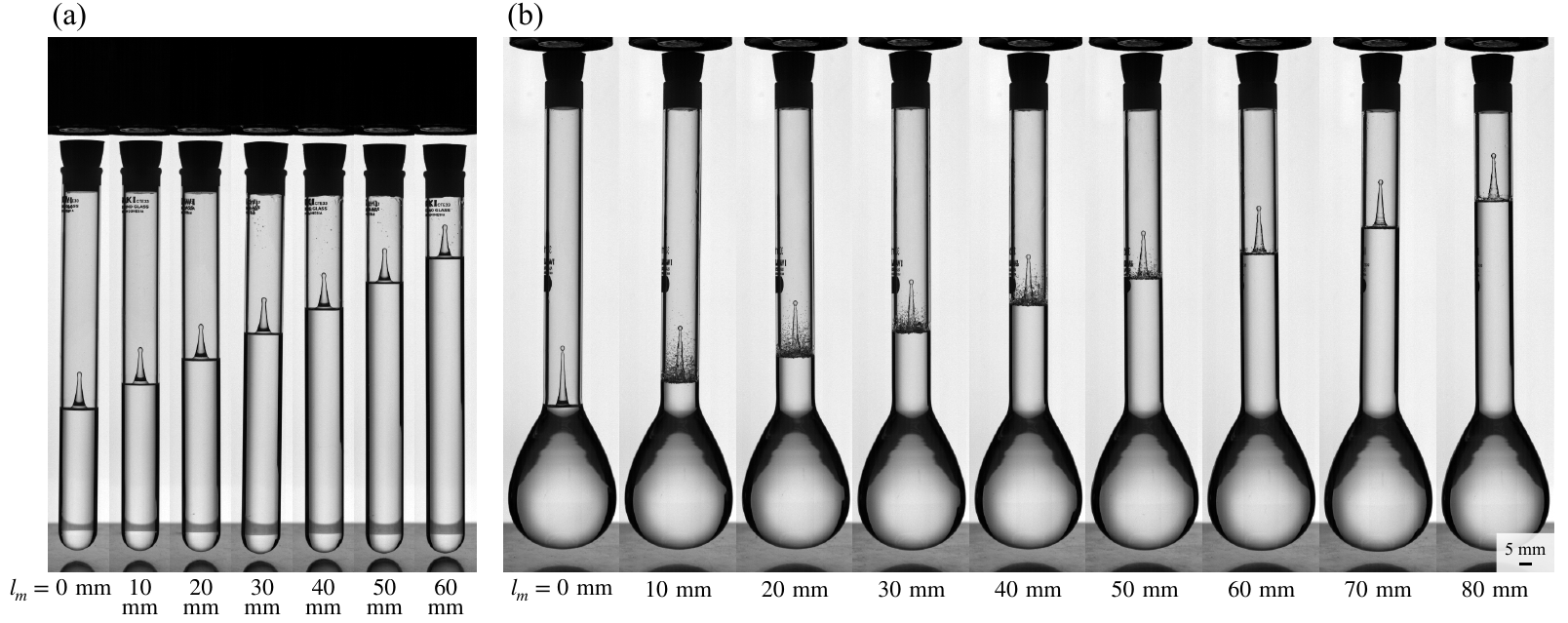}
    \caption{The behavior of impact-induced focused liquid jets when $l_\mathrm{m}$ is varied in (a) the test tube and (b) the Kjeldahl flask. The images were taken 15 ms after container impact, dropped from a height of 6 mm, with $l_\mathrm{m}$ varying in 10 mm steps from 0 to 60 mm for the test tubes and from 0 to 80 mm for the Kjeldahl flasks. Enlarged views of the gas-liquid interface is shown in the upper right. In the Kjeldahl flask, intensive oscillation of the gas-liquid interface was observed, and the oscillation behavior and the accompanying atomization behavior changed with $l_\mathrm{m}$.}
    \label{fig:Jet-lm}
\end{figure}

As a representative example, Fig. \ref{fig:Jet-time} shows the jet generation up to 30 ms after the container impact in a test tube and a Kjeldahl flask filled with ethanol with the condition of $l_\mathrm{m}=0$ mm. The time interval between each image is 5 ms.
The movies of the jet ejection are shown as supplementary material.
Both containers filled with ethanol fall freely from a height of 6 mm when the electromagnet is turned off and impact a metal floor material ($t=0$ ms).
Inside the containers, the flow focuses at the concave gas-liquid interface and a narrow liquid jet is ejected and elongated, results in the formation of droplets ($t=5-30$ ms).
These phenomena are similarly observed in both containers. 
However, comparison of Fig. \ref{fig:Jet-time}(a) and (b) show that the liquid jet extends faster in the Kjeldahl flask than in the test tube.
Fig. \ref{fig:Jet-lm} shows images of the focused liquid jets generated in the test tubes and the Kjeldahl flask at 15 ms after impact of the containers when the height of the liquid column in the middle part $l_\mathrm{m}$, is varied.
In the test tube, the degree of elongation does not change regardless of $l_\mathrm{m}$, whereas in the Kjeldahl flask, the degree of elongation tends to decrease as $l_\mathrm{m}$ increases.

\begin{figure}
    \centering
    \includegraphics[width=0.8\linewidth]{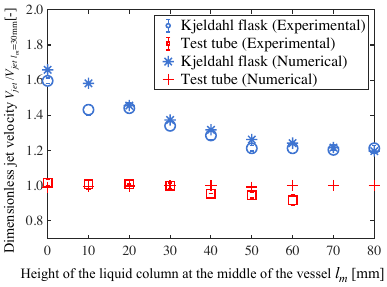}
    \caption{Normalized jet velocity for varying the liquid height in the middle of containers $l_\mathrm{m}$. The experimentally obtained jet velocities $V_\mathrm{jet}$ and the numerically calculated velocity at the bottom of the gas-liquid interface $V$ in both containers are normalized to the velocity at $l_\mathrm{m}=30$ mm in the test tube, respectively. Each experimental plot is the mean of three trials, and the error bars represent the standard deviation.}
    \label{fig:Vjet-all}
\end{figure}

The jet velocities $V_\mathrm{jet}$ are compared in Fig. \ref{fig:Vjet-all} for both the test tube and the Kjeldahl flask as liquid column height $l_\mathrm{m}$ in the middle of containers is varied.
The jet velocities $V_\mathrm{jet}$ in both containers are normalized to the velocity at $l_\mathrm{m}=30$ mm in the test tube, respectively.
Comparing the normalized jet velocities of the two containers, the Kjeldahl flask produce a faster jet than the test tube for all $l_\mathrm{m}$ conditions.
The difference in jet velocity was greatest at $l_\mathrm{m}=0$ mm, where the jet velocity in the Kjeldahl flask is approximately 1.6 times greater than that in the test tube.
The Kjeldahl flask has a maximum $V_\mathrm{jet}$ at $l_\mathrm{m}=0$ mm, and as $l_\mathrm{m}$ increases, the $V_\mathrm{jet}$ decreases and approaches the $V_\mathrm{jet}$ of the test tube, while the jet velocity in the test tube is almost constant regardless of $l_\mathrm{m}$, as also observed in Fig.\ref{fig:Jet-lm}.
The tendency for the jet velocity to be nearly constant regardless of the liquid height in the test tube at the same container impact velocity, unless cavitation occurs, was also shown in previous studies (\citet{Kiyama2014-bh, Kiyama2016-tf}).
This indicate that the mass of the whole container-liquid system has almost no effect on the jet velocity at the same container impact velocity.
In addition, the newly identified trend of a decrease in $V_\mathrm{jet}$ with increasing $l_\mathrm{m}$ in the Kjeldahl flask also supports that the jet did not increase in speed with increasing mass in the Kjeldahl flask.
Since the liquid velocity obtained by impact is determined by the pressure impulse field as described in Chapter \ref{Theory}, the mechanism of jet acceleration in Kjeldahl flasks is clarified by investigating the pressure impulse field in the both containers in \ref{Results-Sub2}.

Although it is not the primary interest of this study, here we report the interesting atomization phenomena observed in the experiment (see Fig.\ref{fig:Jet-lm}).
In the Kjeldahl flask, a significant oscillation of the gas-liquid interface was observed immediately after the container impact, and the concave shape gas-liquid interface was focused while oscillating.
Vibration was also observed during jet elongation at the bottom of the gas-liquid interface.
In the test tube, the jet behavior was similar regardless of $l_\mathrm{m}$, while in the Kjeldahl flask, the behavior of the oscillation of the gas-liquid interface and the intensity of atomization changed depending on $l_\mathrm{m}$.
In particular, at $l_\mathrm{m}=10$ mm, the gas-liquid interface oscillated violently and small droplet scattering was observed, which was not apparent at $l_\mathrm{m}=0$ mm.
These phenomena of small droplet formation at the gas-liquid interface have been observed when a liquid container (\citet{Hashimoto1980-ys, Hashimoto1987-hh}), a plate with a droplet placed (\citet{James2003-vv, Kustron2023-qv}), a nozzle with a droplet attached to its tip (\citet{Yuan2022-hv}) were vibrated in the vertical direction.
The similarity with the previous study suggests that the atomization phenomenon observed in this study is caused by the vibration of the liquid column in the container.

\subsection{Comparison with a numerical solution}
\label{Results-Sub2}
In this section, we investigate the mechanism of jet acceleration in Kjeldahl flasks by solving the Laplace equation numerically in the axial symmetric cylindrical coordinate system to obtain the pressure impulse distribution in both containers used in the experiment.
We solved the Laplace equation numerically with the boundary conditions described in \ref{Methods-Sub2} and obtained pressure impulse fields.
Here, we show the $l_\mathrm{m}=0$ mm condition as a representative example in Fig. \ref{fig:Pi-field}(a), where the largest difference of the jet velocity between containers was observed.
The white lines are the iso-pressure impulse lines, which are drawn at intervals of 0.4 kg/($\mathrm{m\cdot s}$).
From the obtained pressure impulse field in Fig. \ref{fig:Pi-field}(a) and Eq.\ref{Eq:Laplace}, we calculated the liquid velocity at the bottom of the gas-liquid interface and compared with the experimentally measured jet velocity.
The numerically obtained liquid velocities at the bottom of the gas-liquid interface in both containers with varying $l_\mathrm{m}$ were normalized to the velocity at $l_\mathrm{m}=30$ mm in the test tube, respectively.
We show the normalized interfacial velocities in Fig. \ref{fig:Vjet-all} and compare with the experimentally obtained normalized jet velocities.
The results show that the maximum velocity increase of the Kjeldahl flask relative to the test tube is observed at $l_\mathrm{m}=0$ mm both in the numerical results and in the experiment, and the degree of increase is in good agreement.
In addition, the deceleration trend of the jet velocity with increasing liquid height $l_\mathrm{m}$ in the Kjeldahl flask observed in the experiment was also well reproduced by the numerical results.
Although in the numerical solution, we merely calculate the interfacial velocity from the gradient at the bottom of the gas-liquid interface, it reproduces the trend of the jet velocity observed in the experiment well.
This indicates that the jet velocity trend observed in this study is not due to flow focusing at the gas-liquid interface, but the inertia of liquid especially at the central axis of containers.
These results confirm that the present boundary conditions based on the pressure impulse theory can well explain the experimentally observed phenomena.
In particular, the pressure impulse distribution on the central axis of containers is expected to reveal the mechanism of jet velocity increase in the Kjeldahl flask.

On the basis of the numerically obtained pressure impulse field, the mechanism of jet velocity acceleration in a Kjeldahl flask is discussed.
Examining the pressure impulse on the central axis of the container in Fig. \ref{fig:Pi-field}(a), it was confirmed that the pressure impulse is nearly linear with $z$ in the test tube, while it is  non-linear in the Kjeldahl flask (see Fig. \ref{fig:Pi-field}(b)).
This non-linearity also results in a larger pressure impulse gradient at the gas-liquid interface ($z=56$ mm) than in the test tube, causing faster liquid jet generation.
Comparing the pressure impulse distribution on the central axis of the Kjeldahl flask with that of the test tube, the most noticeable difference is at the converging part of the container where the cross-sectional area of the container decreases between the neck and the bottom of the container ($z=20-56$ mm).
Therefore, we expect the relationship between the converging shape of the container and the non-linearity of the pressure impulse distribution.

Here, we consider a simplified shape of the converging part of a Kjeldahl flask and investigate the effect of the degree of convergence on the pressure impulse distribution in \ref{Results-Sub3}.
The converging part of the the Kjeldahl flasks (Fig. \ref{fig:Pi-field}, $z=20-56$ mm) are generally uniformly converging, and the cross-sectional area decreases monotonically.
Although the concave-shaped deformation of the gas-liquid interface has a significant effect on the pressure gradient just below the gas-liquid interface (\citet{Antkowiak2007-ax}), the effect on the linearity of the pressure impulse distribution over the entire container is negligible in long enough containers, as indicated in Fig. \ref{fig:Pi-field}(b).
In addition, at the cross section on the border between two boundary conditions (Fig. \ref{fig:Pi-field}, $z=20$ mm), the equal pressure impulse line is mostly parallel to the r-axis, which means that the pressure impulse value is roughly uniform with the value given at the wall.
Therefore, we consider a simple monotonically converging-shaped container that simulates the converging region of a Kjeldahl flask, consisting of a flat gas-liquid interface, a flat container bottom and a straight container wall.
By changing only the degree of convergence, we clarify the relationship between the degree of convergence of the container, the pressure impulse distribution, and the liquid velocity at the gas-liquid interface in \ref{Results-Sub3}.

\subsection{Effect of converging shape of containers}
\label{Results-Sub3}
The previous section showed a relationship between the container's shape and the pressure impulse distribution on the central axis of the container.
In this section, we investigate the effect of the degree of convergence on the non-linearity of the pressure impulse distribution and the resulting change in the gas-liquid interfacial velocity.

\begin{figure}
    \centering
    \includegraphics[width=0.8\linewidth]{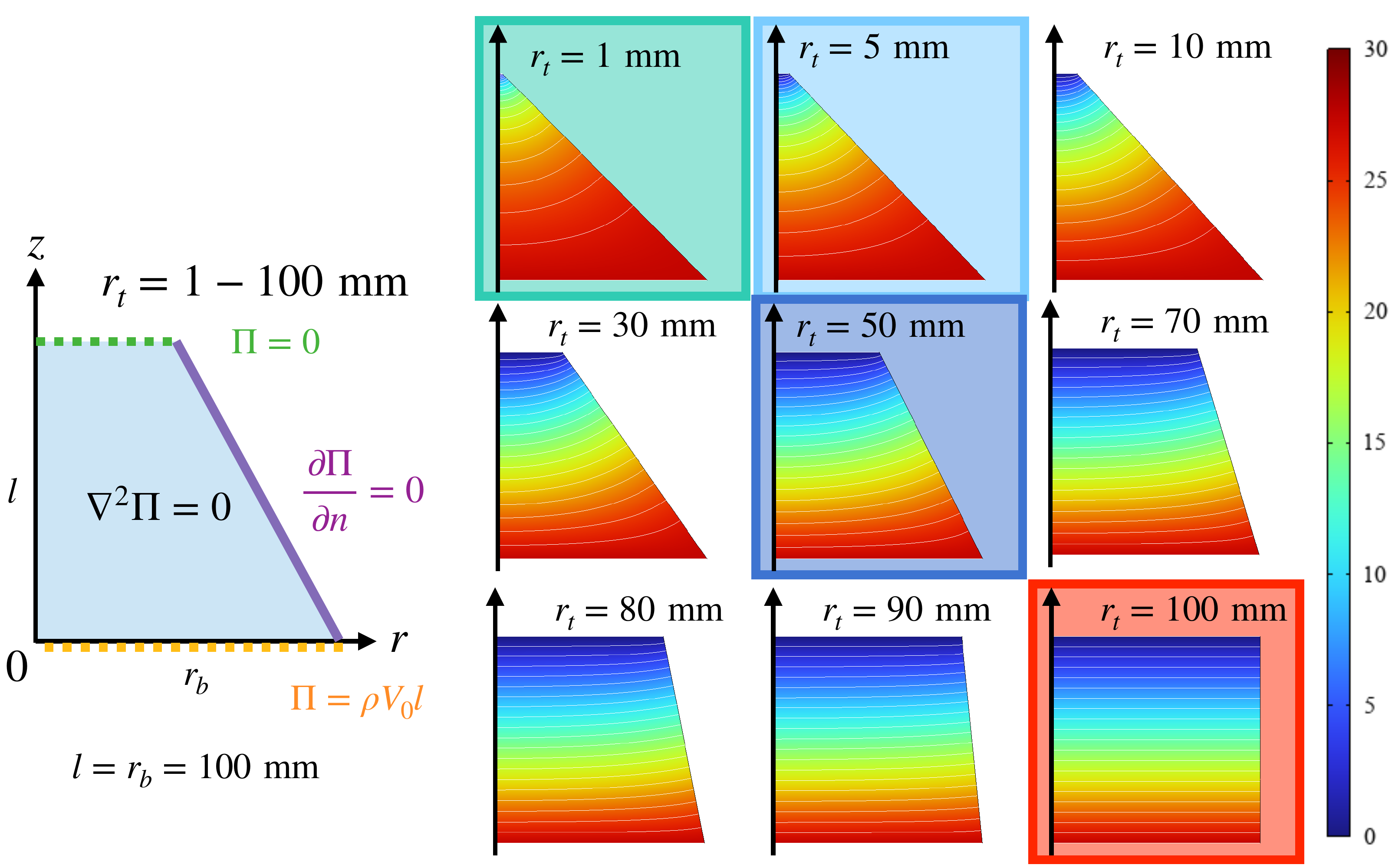}
    \caption{Boundary conditions. Pressure impulse distribution.}
    \label{fig:Pi-field-simple}
\end{figure}

\begin{figure}
    \centering
    \includegraphics[width=0.8\linewidth]{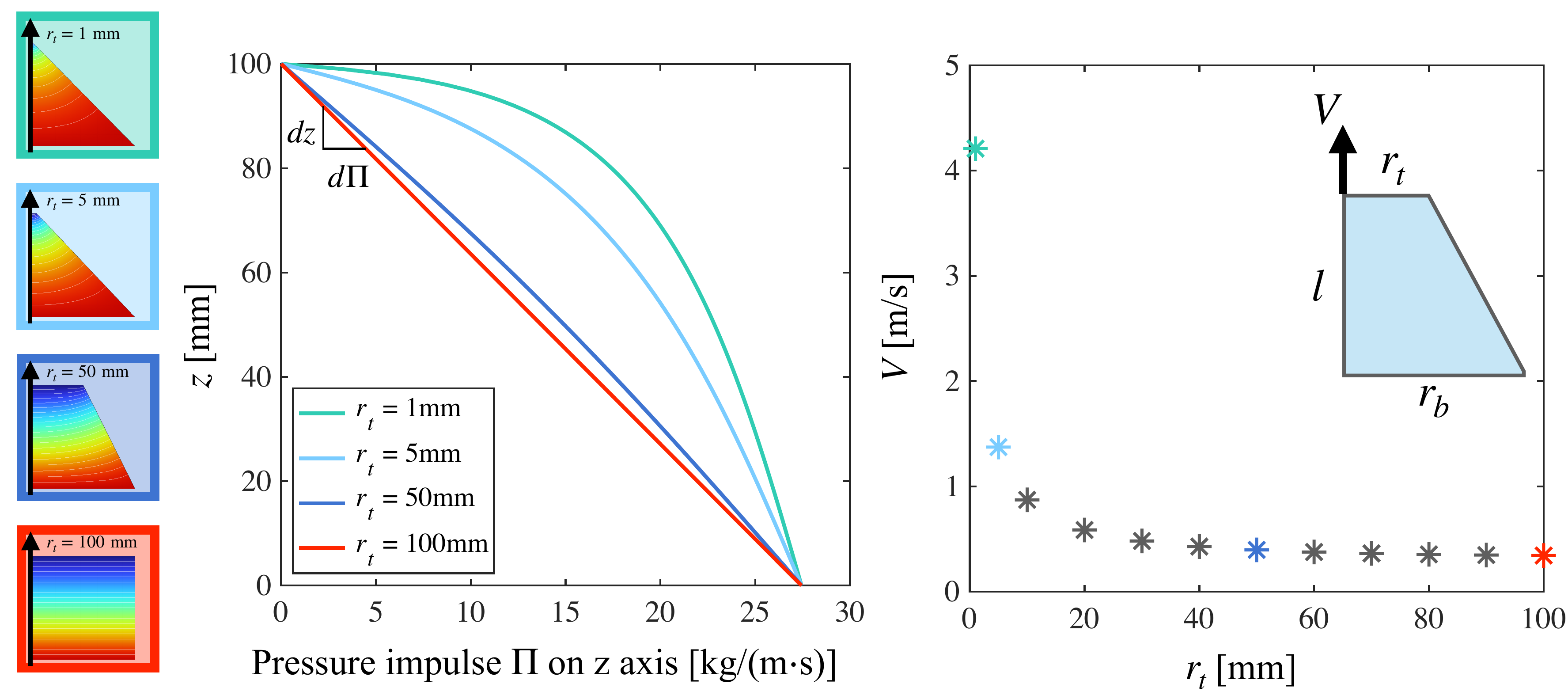}
    \caption{(a) The pressure impulse on the central axis of the container ($z$-axis) for the four conditions of $r_\mathrm{t}= $1, 5, 50, and 100 mm. (b) The velocity at the center of the gas-liquid interface $V$ with varying $r_\mathrm{t}$}
    \label{fig:z-Pi-V-rt}
\end{figure}

To investigate the relationship between the degree of convergence and the nonlinear pressure impulse distribution, numerical analysis is performed on a simple axial symmetric converging container with the boundary conditions shown in Fig. \ref{fig:Pi-field-simple}, which are same as boundary conditions in \ref{Results-Sub2}. 
Among the parameters that determine the container shape, we fixed the liquid height $l$ and the bottom radius $r_\mathrm{b}$ to 100 mm and varied the top radius $r_\mathrm{t}$ from 100 to 1 mm to vary the inclination of the container wall (i.e., the degree of convergence $r_\mathrm{t}/r_\mathrm{b}$).
The pressure impulse distribution in the container is obtained by numerically solving the Laplace equation in axial symmetric cylindrical coordinates under each $r_\mathrm{t}$ condition.
Fig. \ref{fig:Pi-field-simple} shows the pressure impulse fields obtained for each $r_\mathrm{t}$ condition.
The white lines are the iso-pressure impulse lines, which are drawn at intervals of 1.5 kg/($\mathrm{m\cdot s}$).
The pressure impulse on the central axis of the container ($z$-axis) for the four conditions of $r_\mathrm{t}= $1, 5, 50, and 100 mm is shown in Fig. \ref{fig:z-Pi-V-rt}(a), and the velocity at the center of the gas-liquid interface $V$ is shown in Fig. \ref{fig:z-Pi-V-rt}(b).
In the four conditions, $r_\mathrm{t}=$ 100 mm represents a container with no change in cross-sectional area similar to the test tube, $r_\mathrm{t}=$ 50 mm represents a container that has the similar degree of convergence with that of the Kjeldahl flask, and $r_\mathrm{t}=$ 1,5 mm indicates a container with further convergence.

Fig. \ref{fig:z-Pi-V-rt} shows that as $r_\mathrm{t}$ becomes smaller and the container becomes more converged, the non-linearity of the pressure impulse distribution on the central axis of the container becomes stronger.
This causes the increase in the gradient of pressure impulse at the gas-liquid interface, resulting in an increase in the liquid velocity.
For the case of $r_\mathrm{t}=$ 50 mm, where the container has the similar degree of convergence as the Kjeldahl flask, we obtained 1.5 times faster velocity compared to the case of $r_\mathrm{t}=$ 100 mm, which is comparable to the degree of increase for the test tube in the Kjeldahl flask obtained in the experiment.
This indicates the validity of the present verification, in which the walls of the Kjeldahl flask are simulated by linear walls. 
We confirmed that the increase in the pressure impulse gradient at the gas-liquid interface is due to vessel convergence.

Since the Laplace equation is solved so that the continuity equation is satisfied, we expected that the gradient of the pressure impulse changes in order to make the velocity field compatible with the change in the cross-sectional area of the container.
Therefore, we expect a relationship between the change in the cross-sectional area of the plane perpendicular to the center axis of the container and the nonlinear distribution of pressure impulse in the converging-shaped container.
Since the whole system of container and liquid feels acceleration in the vertical direction and obtains velocity change in that direction from the macroscopic point of view, we consider only the cross-sectional area change in the plane perpendicular to the center axis of the container.

\begin{figure}
    \centering
    \includegraphics[width=0.7\linewidth]{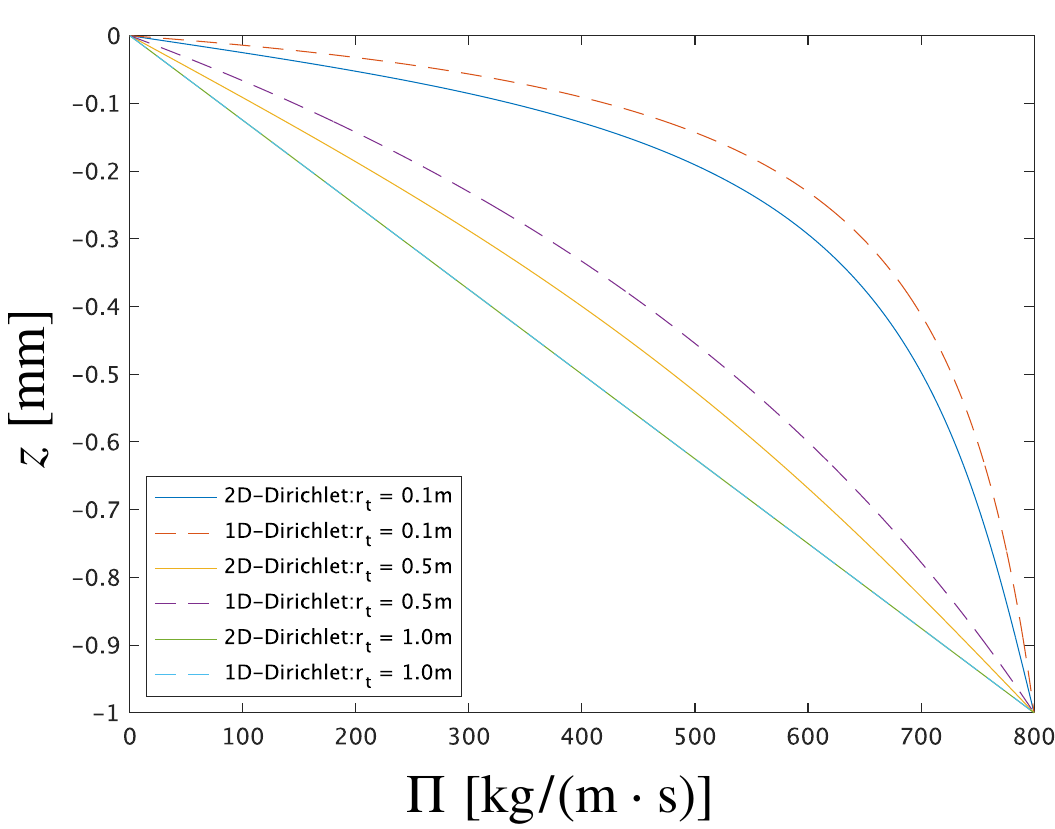}
    \caption{Comparison of the pressure impulse distribution on the center axis of the container $\Pi(z)|_{r=0}$ obtained by solving the one-dimensional Laplace equation analytically and obtained by solving the axial symmetric two-dimensional Laplace equation numerically.}
    \label{fig:z-Pi_2D1D}
\end{figure}

\begin{figure}
    \centering
    \includegraphics[width=0.7\linewidth]{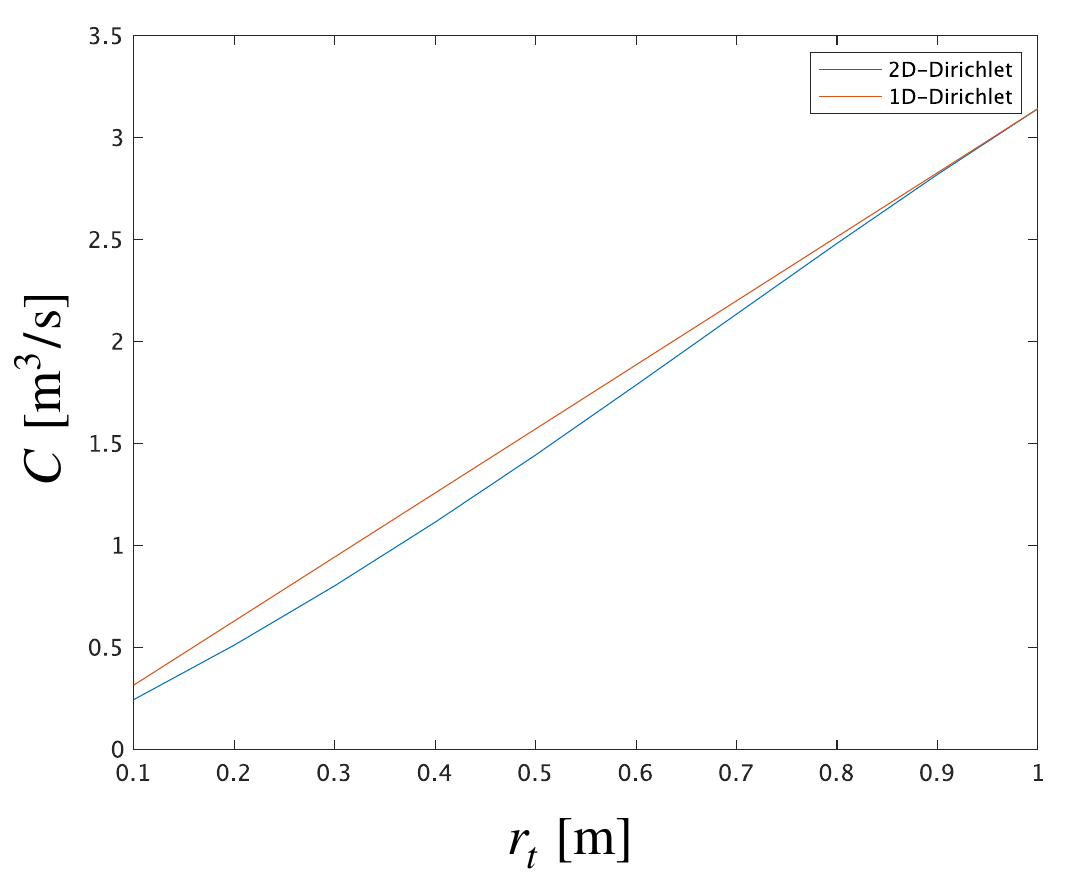}
    \caption{Comparison of the volume flow rate $C$ obtained by solving the one-dimensional Laplace equation analytically and obtained by solving the axial symmetric two-dimensional Laplace equation numerically.}
    \label{fig:C-rt_2D1D}
\end{figure}

Next, to understand the increase in the pressure impulse gradient at the gas-liquid interface associated with the converged shape of the container, we solve the reduced-order model of the pressure impulse distribution on the central axis of the container.
Considering the change in the cross-sectional area perpendicular to the center axis $A(z)$ and the representative velocity on the central axis of the container $V(z)|_{r=0}$, continuous equation can be written as below.

\begin{equation}\label{Eq:continuous}
V(z)|_{r=0} A(z)=C,
\end{equation}
$C$ is the volume flow rate that is constant in each cross-section of the container.
From Eq. \ref{Eq:continuous} and the shortened form of the incompressible Navier-Stokes equation (Eq.\ref{Eq:pressure-impulse}), we can derive

\begin{equation}\label{Eq:general-solution}
\frac{\partial \Pi}{\partial z}= -\frac{\rho C}{A(z)} ,
\end{equation}
We obtained the general solution of the pressure impulse distribution on the container center axis by integrating Eq. \ref{Eq:general-solution} with $z$.
Applying a pressure impulse of 0 at the gas-liquid interface and the same pressure impulse condition as in Fig.\ref{fig:z-Pi-V-rt} ($\Pi=\rho U_0 l$) at the bottom of the container, a particular solution is obtained as below.
\begin{equation}\label{Eq:theory-Pi}
\Pi(z)|_{r=0}=-\frac{r_{\mathrm{b}}l}{(r_{\mathrm{t}}-r_{\mathrm{b}})z + r_{\mathrm{t}}l} \rho U_0 z,
\end{equation}

\begin{equation}\label{Eq:theory-C}
C =\pi r_{\mathrm{t}} r_{\mathrm{b}} U_0.
\end{equation}
The pressure impulse distribution on the center axis of the container $\Pi(z)|_{r=0}$ and the volume flow rate $C$ obtained by solving the one-dimensional Laplace equation analytically is compared with that obtained by solving the axial symmetric two-dimensional Laplace equation numerically in Fig. \ref{fig:z-Pi_2D1D} and Fig. \ref{fig:C-rt_2D1D} respectively.

Fig. \ref{fig:z-Pi_2D1D} shows the pressure impulse distribution on the center axis of the container $\Pi(z)|_{r=0}$ by solving the one-dimensional Laplace equation analytically and by solving the axial symmetric two-dimensional Laplace equation numerically. Fig. \ref{fig:z-Pi_2D1D} shows a general agreement of two methods in the tendency that the non-linearity of the pressure impulse distribution becomes stronger as the degree of convergence increases.
The result is in perfect agreement when there is no cross-sectional area change.
Similarly, for the volume flow $C$, the value obtained by analytically solving the one-dimensional Laplace equation is compared with the value obtained by numerically solving the axial symmetric two-dimensional Laplace equation.
In numerical analysis, $C$ was obtained by cutting concentric rings at 1 mm intervals across the bottom of the vessel and integrating the volume flow at each ring.
As shown in Fig. \ref{fig:C-rt_2D1D}, the tendency for the volume flow $C$ to decrease with stronger convergence of the container is similar between the two.
This trend may seem surprising to some and will be discussed later in this section.
From these results, the effect of the converging shape on the pressure impulse distribution on the central axis is well understood even if only the one-dimensional cross-sectional area change is considered, without considering the two-dimensional liquid motion. 
The motion of the fluid in a converging-shaped container is found to be governed by the coupling between the change in volume flow and the velocity-increasing effect due to the one-dimensional cross-sectional area change, rather than by the two-dimensional liquid motion.
In summary, the narrower the container, the more the pressure impulse gradient increases due to the change in cross sectional area in one dimension, the more the pressure impulse gradient on the container center axis increases, which in turn increases the pressure impulse gradient at the gas-liquid interface, resulting in a larger gas-liquid interfacial velocity.

\begin{figure}
    \centering
    \includegraphics[width=1.0\linewidth]{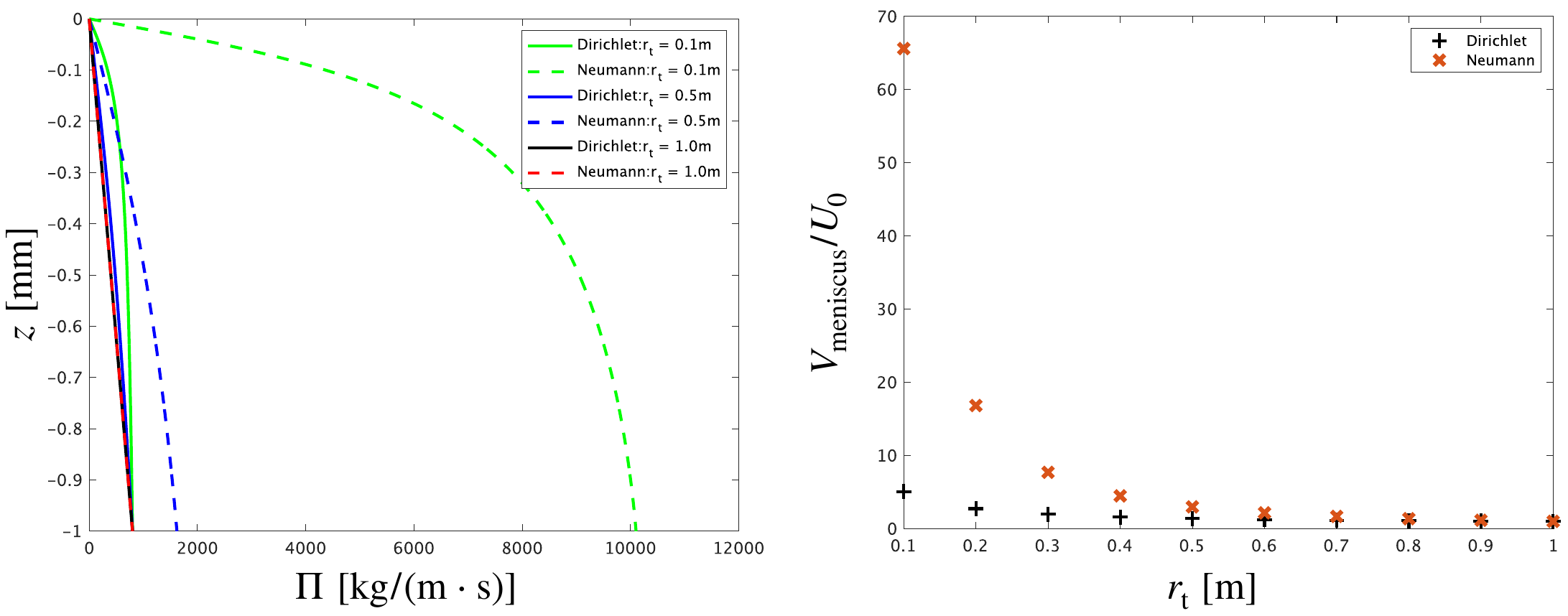}
    \caption{Pressure impulse distribution on the axis, the velocity at the bottom of the meniscus.}
    \label{fig:z-Pi-V-rt_DN}
\end{figure}

It may seem surprising that the volume flow $C$ tends to decrease as the container becomes more converging, even though the drop height is the same and the resulting impact velocity is the same.
Since the same impact velocity is given to the system, it would seem correct to assume that the liquid at the bottom of the container is always given the same velocity change with the container, regardless of the shape of the container, and that the flow rate will be constant.
Therefore, we further consider the variation of the flow rate depending on the degree of convergence.
At the bottom of the same converging-shaped container, we applied the Neumann condition, which gives the constant liquid velocity, and compared it with the Dirichlet condition, which gives the constant pressure impulse.
As shown in Fig. \ref{fig:z-Pi-V-rt_DN}, the Neumann condition, which gives the constant flow rate for all $r_\mathrm{t}$ conditions, does not match the Dirichlet condition, which is consistent with the experimental results described above.
As shown in Fig. \ref{fig:z-Pi-V-rt_DN}, in containers with no cross-sectional area change, both conditions show no difference, so the type of boundary condition was not an issue. (The difference only appears when applied to a container with a converging shap.)
Considering the Kjeldahl flask and if the flow rate at the bottom is constant, the gas-liquid interfacial velocity remains constant regardless of the change in $l_\mathrm{m}$, which does not explain the decrease in jet velocity with increasing $l_\mathrm{m}$ observed in the experiment.
The Dirichlet condition, rather than the Neumann condition, explains the experimental trend better, confirming that the liquid motion associated with impact is not driven by the effect of volume flow at the liquid bottom boundary, but by the pressure impulse generated at the bottom of the container.

We consider the limits of application of the Dirichlet boundary condition, which gives the value of the pressure impulse at the bottom of the container.
The boundary condition that gives the pressure impulse at the bottom of the container sufficiently distant from the gas-liquid interface was based on the assumption that the pressure impulse gradient caused by the deformation of the gas-liquid interface does not significantly affect the pressure impulse distribution over the entire container.
Therefore, in situations where the distance between the gas-liquid interface and the bottom of the container is too close, (i.e., when the total length of the liquid is not sufficiently large relative to the depth of the gas-liquid interface,) the Dirichlet boundary condition may not capture the real phenomena.

\section{Conclusion}
\label{Conclusion}
In order to investigate the effect of the converging container shap on the gas-liquid interfacial behavior in a container experiencing a sudden velocity change, this paper compares the jet velocities in a conventional test tube with no cross-sectional area change and a Kjeldahl flask with a cross-sectional area change, and performs an experiment to generate an impact-induced focused liquid jet.
The results showed that the Kjeldahl flask produced a faster jet than the test tube, even though the impact velocity was the same in both containers. 
In addition, the jet velocity in the Kjeldahl flask varied with the height of the liquid, a trend not seen in previous jet formation in test tubes.
We investigated the mechanism of jet acceleration by numerically solving the Laplace equation in an axial symmetric cylindrical coordinate system based on the pressure impulse theory.
By using the Dirichlet boundary condition, which gives the constant pressure impulse value itself as the boundary condition at the bottom of the container when solving Laplace's equation, we were able to reproduce the experimental difference in jet velocity between the two containers and the relationship between the liquid height and jet velocity.
We confirmed that the impact-induced focused liquid jet is driven by the pressure impulse at the bottom of the container.
In addition, the pressure impulse field obtained by the numerical solution suggests a relationship between the container's converging shape and the pressure impulse gradient along the center axis of the container.

Numerical analysis was also performed for a simple container with the constant inclined side wall, a flat gas-liquid interface, and a flat bottom, varying the inclination of the container walls and the degree of convergence.
The results show that the non-linearity of the pressure impulse distribution along the central axis of the container increases with container convergence, which in turn increases the pressure impulse gradient at the gas-liquid interface and the liquid velocity.

To confirm the effect of the change in the cross-sectional area of the container, we analytically solved a reduced-order model, and obtained the same trend as the numerical solution of the axial symmetric two-dimensional Laplace equation. 
We conclude that the change in flow rate and pressure impulse gradient near the gas-liquid interface due to the change in cross-sectional area results in a change in the gas-liquid interfacial velocity.
These results indicate that the use of a container with more converged shap can increase the jet velocity at the same impact velocity, which has potential engineering applications as a non-diluted ejection and application technology for high-viscosity liquids such as paints and adhesives in terms of increasing the ejection viscosity of the liquid.
In addition, it is useful for the development of needle-free injection technology using impact-induced focused liquid jets.
In addition to the medical and engineering applications of jet velocity control, the system has important implications for the numerical and analytical study of rapid changes in various liquid motions, not limited to jets.
For example, it may be used to understand the motion of liquid fuel in a storage tank when a space rocket suddenly accelerates or stops(Baumbach et al. 2005!, Hopfinger and Baumbach 2009!)

\section{Acknowledgements}
\label{Acknowledgements}

This work was supported by the National Institutes of Health [grant numbers xxxx, yyyy]; the Bill Melinda Gates Foundation, Seattle, WA [grant number zzzz]; and the United States Institutes of Peace [grant number aaaa].




\bibliographystyle{elsarticle-num-names} 
\bibliography{Watanabe_IJMF_2025_0205.bib}







\end{document}